\documentclass[sigconf, authorversion]{acmart} 

\AtBeginDocument{%
  \providecommand\BibTeX{{%
    \normalfont B\kern-0.5em{\scshape i\kern-0.25em b}\kern-0.8em\TeX}}}

\setcopyright{acmcopyright}
\copyrightyear{2024}
\acmYear{2024}
\acmDOI{10.1145/3649902.3653942}

\acmConference[ETRA '24]{2024 Symposium on Eye Tracking Research and Applications}{June 4--7, 2024}{Glasgow, United Kingdom}
\acmBooktitle{2024 Symposium on Eye Tracking Research and Applications (ETRA '24), June 4--7, 2024, Glasgow, United Kingdom}
\acmISBN{979-8-4007-0607-3/24/06}

\begin{document}

\title[Exploring Communication Dynamics]{Exploring Communication Dynamics: Eye-tracking Analysis in Pair Programming of Computer Science Education}

\author{Wunmin Jang}
\authornotemark[1]
\email{wunmin.jang@tum.de}
\orcid{0009-0005-2288-6732}
\affiliation{%
  \institution{Technical University of Munich}
  \streetaddress{Arcisstraße 21}
  \city{Munich}
  \country{Germany}
  \postcode{80333}
}

\author{Hong Gao}
\email{hong.gao@tum.de}
\orcid{0000-0003-3934-433X}
\affiliation{%
  \institution{Technical University of Munich}
  \streetaddress{Arcisstraße 21}
  \city{Munich}
  \country{Germany}
  \postcode{80333}
}

\author{Tilman Michaeli}
\email{tilman.michaeli@tum.de}
\orcid{0000-0002-5453-8581}
\affiliation{%
  \institution{Technical University of Munich}
  \streetaddress{Arcisstraße 21}
  \city{Munich}
  \country{Germany}
  \postcode{80333}
}

\author{Enkelejda Kasneci}
\email{enkelejda.kasneci@tum.de}
\orcid{0000-0003-3146-4484}
\affiliation{%
  \institution{Technical University of Munich}
  \streetaddress{Arcisstraße 21}
  \city{Munich}
  \country{Germany}
  \postcode{80333}
}

\renewcommand{\shortauthors}{Jang, et al.}

\begin{abstract}
Pair programming is widely recognized as an effective educational tool in computer science that promotes collaborative learning and mirrors real-world work dynamics. However, communication breakdowns within pairs significantly challenge this learning process. In this study, we use eye-tracking data recorded during pair programming sessions to study communication dynamics between various pair programming roles across different student, expert, and mixed group cohorts containing 19 participants. By combining eye-tracking data analysis with focus group interviews and questionnaires, we provide insights into communication's multifaceted nature in pair programming. Our findings highlight distinct eye-tracking patterns indicating changes in communication skills across group compositions, with participants prioritizing code exploration over communication, especially during challenging tasks. Further, students showed a preference for pairing with experts, emphasizing the importance of understanding group formation in pair programming scenarios. These insights emphasize the importance of understanding group dynamics and enhancing communication skills through pair programming for successful outcomes in computer science education.
\end{abstract}




\maketitle

\label{intro}
\section{Introduction}
Pair programming (PP) is a collaborative method where two programmers jointly work on the same tasks, such as design, review, and debugging, effectively boosting productivity and software quality~\cite{williams2000all}. This method requires coordination and shared understanding between programmers~\cite{rimolsronning2022eye}. In computer science (CS) education, educators increasingly explore PP in classroom settings, both formally and informally~\cite{carver2007increased}, to enhance student learning experiences. It is recognized as a highly effective method for promoting learning, teamwork, and the development of communication skills and cooperative competencies~\cite{succi2001extreme, williams2003pair, williams2000strengthening}. This collaborative educational approach encourages active discussion engagement, fosters effective problem-solving, and contributes significantly to communication competency~\cite{bailey2017value}. However, particularly evident among novices, communication challenges frequently pose a notable barrier to the successful outcomes of PP exercises~\cite{begel2008pair, cockburn2000costs}. Therefore, understanding how these pairs communicate and identifying points of breakdown in communication is a valuable avenue for research~\cite{williams2003pair}. Despite the recognized significance of communication in the learning environment, there is limited research on communication skills within CS education.

Incorporating eye-tracking methods into studies on collaborative tasks has yielded fresh insights into collaboration mechanisms~\cite{d2017improving, gupta2016you, stein2004another}. Communication barriers often arise in PP, posing challenges, especially for novices~\cite{begel2008pair, cockburn2000costs}. Therefore, effective communication between pairs is crucial in PP, as silence may signal underlying issues, underscoring communication's vital role~\cite{begel2008pair, murphy2010pair, sanders2002student}. In this sense, our experiments shed light on communication dynamics during PP in CS education, analyzing eye-tracking data with questionnaires and FGIs through triangulation analysis to provide multifaceted insights.

\label{background}
\section{Related Work}

\subsection{Pair Programming}

In CS education, educators are increasingly exploring pair programming approaches in the classroom, both formally and informally ~\cite{carver2007increased}. PP has been associated with higher student performance~\cite{mcdowell2002effects,mcdowell2006pair,nagappan2003improving,williams2002support}, improved confidence ~\cite{mcdowell2006pair}, enjoyment, and satisfaction ~\cite{williams2000strengthening}, and yielding a range of additional advantages, such as enhanced self-sufficiency ~\cite{nagappan2003improving,williams2002support}. Specifically, Lui et al.~\cite{lui2006pair} found that PP proficiently helps programmers solve unfamiliar tasks by investigating how programmers with different abilities offer different results. In debugging, experts spend more time comprehending the program and guiding novices, employing effective strategies such as viewing code as chunks ~\cite{lister2011concrete}. In contrast, students may compare their codes with others and often favor forward reasoning ~\cite{katz1987debugging}. 

In communication paradigms, experts focus on the code and their approach to solving the problem, while novices try to communicate that idea to the expert with a different problem-solving approach but are unsuccessful immediately~\cite{plonka2015knowledge}. Further, increasing research has been utilizing eye-tracking to explore collaborative learning scenarios in PP ~\cite{olsen2015dual, jermann2011collaborative, pietinen2008method, poole2006eye}. It underscores the importance of unobtrusively recording eye-related metrics to mimic natural settings, allowing freedom of movement to prevent cognitive load ~\cite{pietinen2008method, rimolsronning2022eye}. Baheti et al. ~\cite{baheti2002exploring} concluded that the co-location factor is negligible, as spatially distributed pairs benefit similarly, particularly in synchronous and collaborative tasks conducive to shared gaze awareness. These studies demonstrate that eye-tracking is a tool to predict collaboration and comprehension, while unanswered questions remain regarding the influence of pair dynamics on the success of collaboration between pairs, and certain potential indicators that could impact collaboration success remain to be thoroughly investigated ~\cite{villamor2018predicting}.

\subsection{Eye-tracking in Education and Communication Studies}

Eye-tracking studies have expanded into science education, investigating students' comprehension of complex concepts ~\cite{cullipher2015atoms, williamson2013identifying}. Eye movement observations serve as a crucial physiological index for quantifying decision-making processes ~\cite{kurimori1995evaluation,appel2019predicting,bozkir2019assessment,castner2018scanpath,byrne2023exploring}, providing valuable insights into cognitive processing ~\cite{holmqvist2011eye, just1976eyefixations}. Recent studies have demonstrated that utilizing eye-tracking data to provide feedback during collaborative programming tasks enhances collaboration quality and effectiveness in conveying source code locations ~\cite{d2017improving, schneider2017real}. Furthermore, integrating eye-tracking data into feedback systems during collaborative tasks consistently yields significant benefits ~\cite{d2017improving, gupta2016you, stein2004another}.

In PP context, communication barriers often surface, posing challenges for novices ~\cite{begel2008pair, cockburn2000costs}. Effective communication between pairs is emphasized as crucial, as silence may signal potential issues, underlining the pivotal role of communication in PP success ~\cite{begel2008pair, murphy2010pair, sanders2002student}. Pairs experiencing unsuccessful outcomes often express feelings of guilt, frustration, and a sense of time wastage \cite{thomas2003code}. Additionally, mismatches in pair compatibility, particularly when one partner has a lower skill level, can adversely impact satisfaction and performance ~\cite{thomas2003code}. Some students struggle with pairing dynamics and exhibit reluctance to trust their partners' codes in classroom settings [28]. Approximately 50\% of the students cited various difficulties within pairs as hindering communication, a central challenge in the PP process. Moreover, the majority of students expressed a preference for PP over extreme programming ~\cite{sanders2002student}.

\label{methods}
\section{Method}

\subsection{Participants}

For our study we recruited 19 participants (12 male, 7 female). The participants included nine students with Java programming experience and ten experts with over three semesters of programming experience. Experts reported having computer science experience for more than 5 years (80\%), and 3-5 years (20\%), while among students, they had 1-2 years of experience (66.67\%)  and 3-5 years of experience (33.33\%). The experiment comprised two randomly assigned sessions: one student group and expert group, and one mixed group.

\subsection{Study Design}

Our study comprised two sessions: one involving student-only and expert-only groups and another involving mixed (student-expert) groups. Each session consisted of two paired groups, forming two pairs. We employed three data collection methods: eye-tracking, questionnaires, and Focus Group Interviews (FGIs), utilizing triangulation to enhance data validity~\cite{chaparro2005factors, flick2007designing}. Triangulation, involving the use of mixed methods, serves as multi-method research to increase research credibility; specifically, as many social researchers often rely on a single research method, they may face limitations associated with that method, but it offers an opportunity for enhanced reliability by integrating multiple methods~\cite{bryman2004triangulation}. In other words, we utilized a triangulation methodology by integrating FGIs as a qualitative research method, such as eye-tracking and questionnaires, aiming to supplement the limitations of quantitative methods and offer a comprehensive interpretation of the research problem. Before the experiment, we conducted a pilot test with each of the two experts and students to derive problems and supplements from the experimental process and to design a solid experimental environment. The main study was conducted in February 2024, with recordings and transcriptions capturing real-time participants' thought processes, communication strategies, and skill development perceptions. Eye-tracking data, Zoom communication logs, and task execution in Visual Studio Code were recorded to facilitate iterative validation and assessment of communication skills.

\subsection{Tasks}

\begin{figure}[ht]
    \centering
    \includegraphics[width=.90\linewidth]{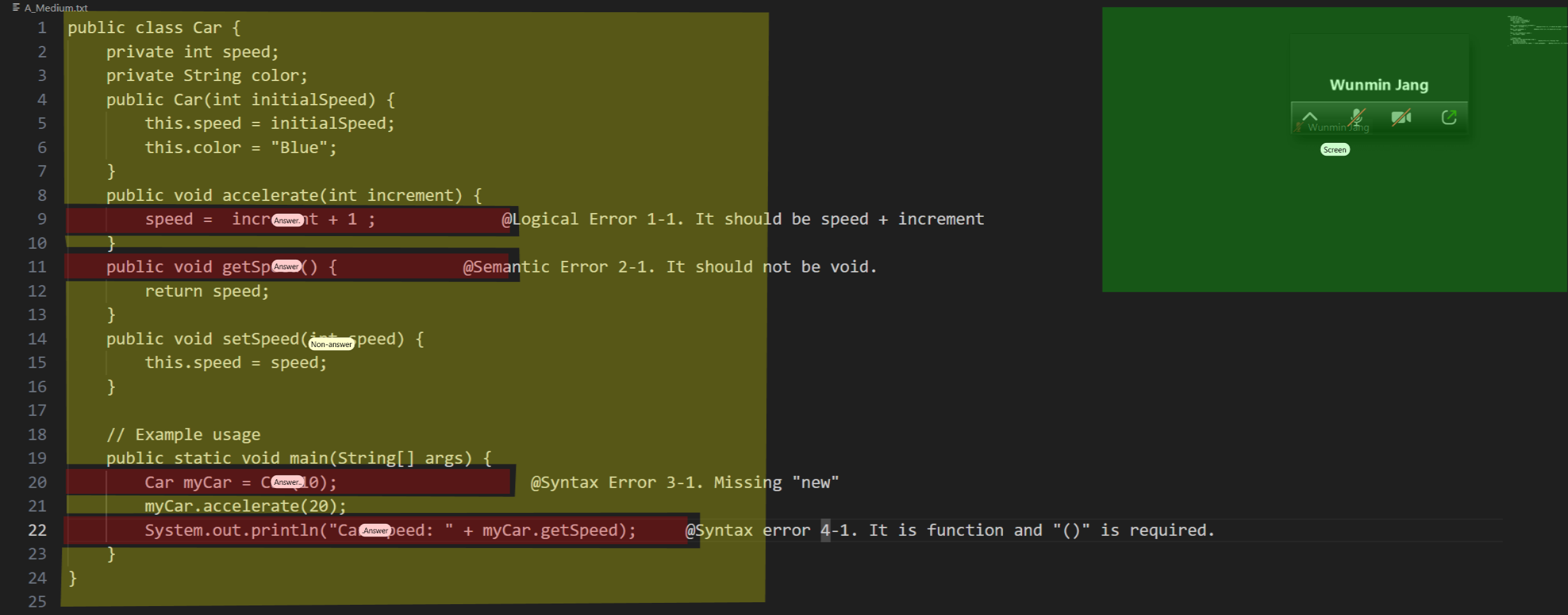}
    \caption{An Example of the Task with the three AOIs displayed: created three large AOIs, which are as follows: Answer, Non-answer, and Screen. This Fig.1 includes the answers with descriptions in the code.}
    \label{fig:enter-label}
\end{figure}

The PP tasks utilized the Java programming language, focusing on debugging tasks to assess performance. Two experts in the computer science field with a Ph.D. and in a doctoral course, respectively, designed the tasks. The tasks were categorized into three difficulty levels: easy, moderate, and hard, each involving identifying syntax, semantic, and logical errors. Task difficulty was assessed through the pilot test, and appropriate adjustments were made accordingly. Success was determined based on the pair's average score compared to the mean. Figure \ref{fig:enter-label} illustrates an example of a debugging task with configured Areas of Interest (AOIs). 

\subsection{Procedure}
Participants provided informed consent before experiments, including procedures like eye-tracking and recording. Pre-questionnaires assessed programming background and demographics. Debugging tasks were conducted for 30 minutes per session, with participants randomly paired across student-student, expert-expert, and expert-student groups. After the first session, participants took a break and were paired with another partner. Communication occurred via Zoom webcam, following PP principles, with one participant as the driver in Visual Studio Code and the other navigating via Visual Studio Live Share. Post-questionnaires included the Conversational Skills Rating Scale (CSRS)~\cite{spitzberg2007csrs} and a Pair Programming Questionnaire~\cite{chaparro2005factors}. FGIs gathered real-time thought processes, communication strategies, and skill development perceptions upon task completion. 

\subsection{Material and Measures}

\subsubsection{Eye-tracking Data}

Eye-tracking data was recorded to collect participants’ eye movements during sessions using Tobii Fusion Pro(\footnote{https://www.tobii.com/products/eye-trackers/screen-based/tobii-pro-fusion}). This screen-based eye tracker has two eye-tracking cameras running at 250 Hz. In the data, AOIs were drawn to map the fixations to the visual stimuli. AOI analysis includes summing all fixations within an area and dividing by the total fixation count in the experiment to calculate the mean fixation duration~\cite{bednarik2004visual}. We created three large AOIs for this study: Answer, Non-answer, and Screen. In AOIs, Answers indicate syntax, semantic, and logical errors, Non-answer is the code parts excluding the Answers, and the Screen demonstrates the Zoom screen segment. We calculated the tracking ratio for data quality. The exclusion criteria was under 70\%, which was driven by investigating the adequacy of research paradigms used in of psychology~\cite{riege2021covert}. During this process, 5 individual data were excluded, and 30 data were included among 35 data and grouped respectively, while every individual's data was stored by session. The mean tracking ratio in the latter group with 30 individual data was 85.0\% (SD = 7.3), compared to 59.3\% (SD = 1.9) in the former group. We investigated participants' fixations and saccades based on expertise, groups, roles, and difficulty levels and visualized them with boxplots.

\subsubsection{Questionnaires}

Participants completed pre-questionnaires for demographic information and programming background, adapted from previous studies. Questions were adjusted by modifying "choices to solve logic problems" to include semantic problems. Programming background was assessed using a test ~\cite{al2016effectiveness}, with modifications, such as replacing "Alice language" with "Java language" and removing irrelevant questions to this research.

After the experiment, participants reported communication and pair programming experience using the CSRS ~\cite{spitzberg2007csrs}, which has been applied to assess communication competency in PP ~\cite{choi2021better}. Modifications were made to the CSRS questionnaire to suit the online PP sessions conducted via Zoom, including clarifications such as "through Zoom screen" for relevant questions regarding non-verbal communication cues. Additionally, a pair programming experience questionnaire developed by Chaparro et al. ~\cite{chaparro2005factors} was utilized. This questionnaire evaluated participants' preferences, learning perceptions, and task difficulty ratings during PP sessions. Some questions were revised to be aligned with the objectives of this study. For example, "8. Lean toward partner (neither too far forward nor too far back)" was deleted, which is a different condition from our experiment that pairs do not sit next to each other. We added "through Zoom screen" to the other questions, such as "11. Use of eye contact" to clarify the questions. At the end of the questionnaires, participants rated the difficulty level of the debugging tasks in order to compare differences in communication skills based on task complexity using a 5-point Likert scale from 1 (Very easy) to 5 (Very difficult).

\subsubsection{Focus Group Interview (FGI)}

Focus Group Interview (FGI) is a qualitative research method. It consists of sets of individuals who share experiences and discuss a topic. It provides deeper and more information than individual interviews ~\cite{thomas1995comparison}. Furthermore, it can generate a large amount of data relatively quickly ~\cite{rabiee2004focus}, and also expect various opinions based on the synergy of the group interaction ~\cite{green2003short}.
Kreuger~\cite{krueger1998moderating} divided questions into different types, consisting of opening, introductory, transition, key, and ending. Based on these phases, we developed the research questions, collected the communication process, and analyzed overall opinions and differences in them. We performed the FGIs two times with students and with experts, respectively. The example questions for FGIs were as follows: \textit{"Could you describe the communication strategies you employed across the roles when you did it based on your partner's positions?", "Are there any challenges or improvements in communication that you experienced across the role and partner's positions during the pair programming sessions?"}

\label{Results}
\section{Results}

\begin{figure*}[ht]
    \centering
    \begin{tabular}{cc}
        \includegraphics[width=0.40\textwidth]{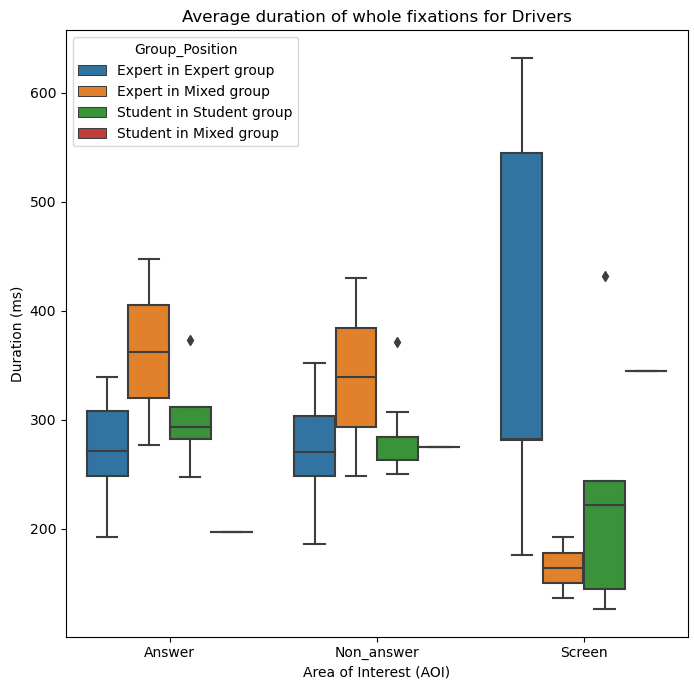} & \includegraphics[width=0.40\textwidth]{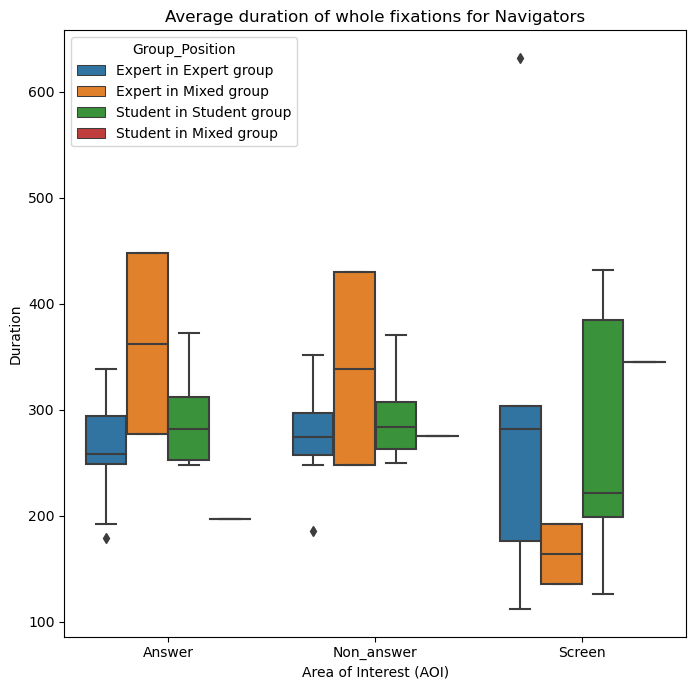} \\
        \includegraphics[width=0.40\textwidth]{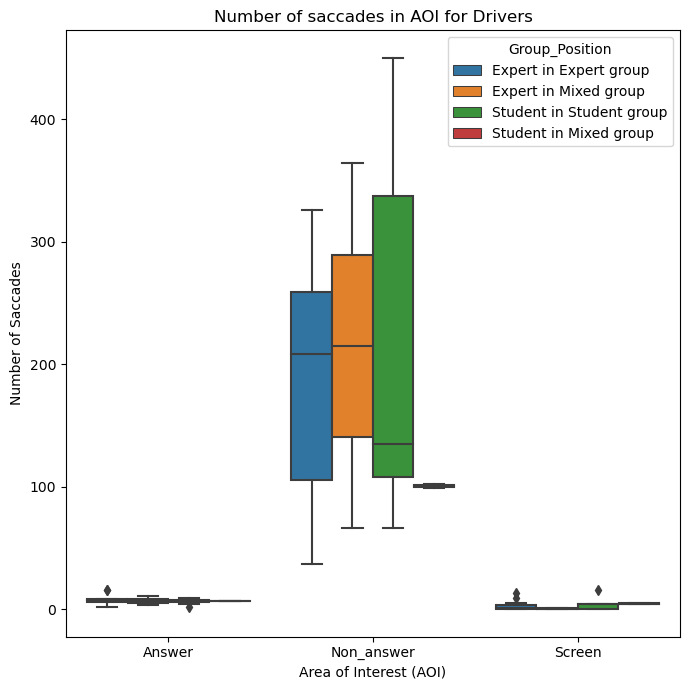} & \includegraphics[width=0.40\textwidth]{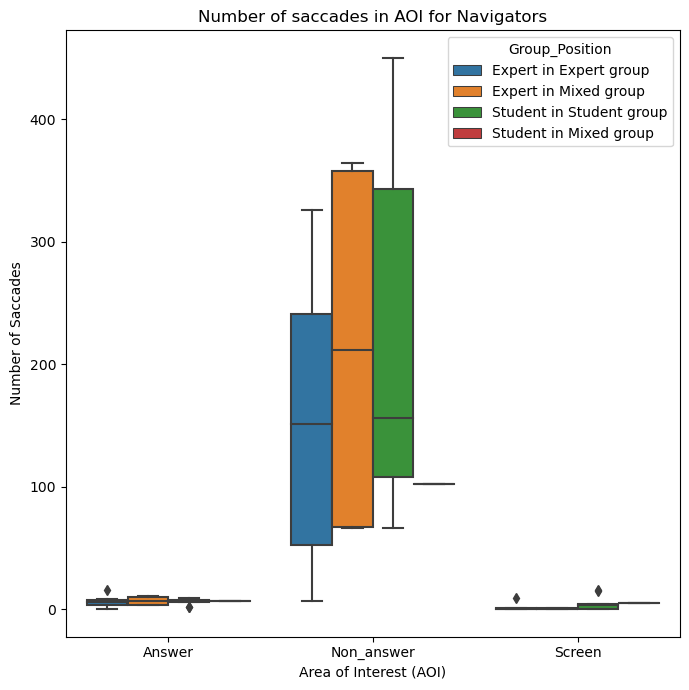} \\
    \end{tabular}
    \caption{Average duration of whole fixations (top row) and number of saccades (bottom row) based on AOIs}
    \label{adfns}
\end{figure*}

\subsection{Eye-Tracking Analyses} 

We performed the Shapiro-Wilk test to verify the normality assumption. As the data did not meet the normality assumption, the Mann-Whitney U test was used for independent samples ~\cite{kim2015t}, and the Kruskal-Wallis~\cite{kruskal1952use} was used to assess the differences among three groups as a nonparametric statistical test. 
Thus, the Kruskal-Wallis test revealed significant differences in the average duration of whole fixations among groups (\textit{H} (2) = 7.489, \textit{p} = .023). Similarly, the Mann-Whitney U test indicated significant differences in the number of saccades between positions (U = 40502.0, \textit{p} = .035) and between expert and student groups (U = 14653.0, \textit{p} = .029). Furthermore, the Kruskal-Wallis test showed notable differences in the number of saccades among groups (\textit{H} (2) = 6.346, \textit{p}= .042). All \textit{p}-values were significant at the $\alpha$
 = 0.05 level in this result.

\begin{figure*}[ht]
    \centering
    \includegraphics[width=0.45\textwidth]{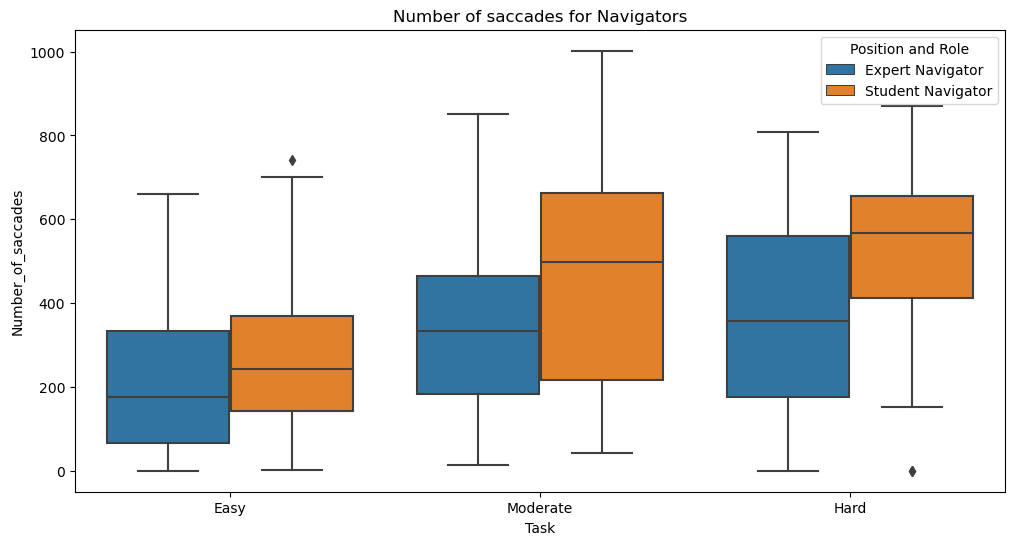}
    \includegraphics[width=0.45\textwidth]{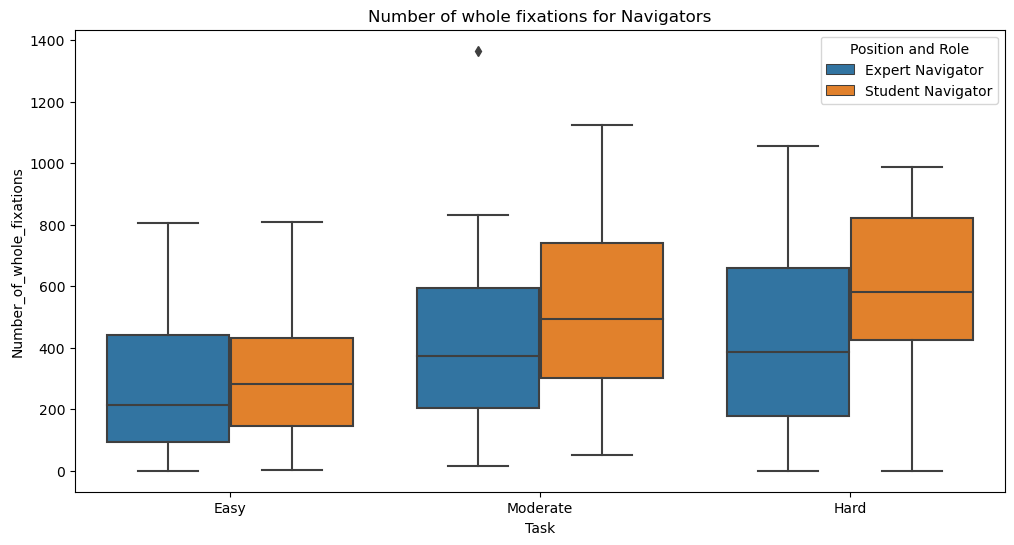}
    \caption{Comparison of number of whole fixations and saccades based on difficulty level}
    \label{tasklevel}
\end{figure*}

We found that experts in the mixed group consistently exhibit the longest fixation duration in both answer and non-answer segments, regardless of their roles (see Figure \ref{adfns}). Figure \ref{adfns} illustrates both an average duration of whole fixations and a number of saccades across various groups and roles, with subfigures depicting fixation duration categorized into different segments, including answer, non-answer, and screen segments. This result confirms that experts devote more time to assisting novices, utilizing effective strategies such as chunking code for better comprehension ~\cite{lister2011concrete}. Conversely, it is the shortest compared to other group positions in the screen segment. Thus, it suggests increased challenges and complexity for experts in mixed groups. However, when experts work with experts of similar expertise, experts show noticeably lower fixation duration than those in mixed groups, particularly in answer and non-answer segments. On the one hand, students in mixed groups demonstrate the most prolonged average duration of whole fixation on screen, regardless of the roles. It can be assumed that they prefer to communicate with experts rather than focus on the tasks. Furthermore, students generally have shorter fixation duration than experts in answer and non-answer segments, especially when paired with experts, which is notably low. This finding also corresponds with the fact that novice programmers risk disengagement in PP sessions, especially when they are paired with an expert~\cite{plonka2012disengagement}. It can be inferred that these differences in fixation duration highlight the impact of partners' programming backgrounds in PP and influence participants' actions and communication during sessions. 

Given that the non-answer part occupies most of the codes, the number of saccades is focused on the non-answer segment. It can be seen that participants tend to focus on the codes rather than communicating with their partners. However, students in mixed groups have a remarkably low number of saccades compared to the other groups, although in the non-answer part.

Figure \ref{tasklevel} depicts the number of whole fixations and saccades for navigators according to task difficulty levels. They demonstrate that as tasks become more challenging, there is an increase in both the number of whole fixations and saccades. Additionally, students generally exhibit higher numbers of whole fixations and saccades compared to experts. This suggests that when navigators encounter difficult tasks, they require increasing cognitive effort and face heightened communication demands. It implies potential communication challenges for students during the sessions, whereas communication appears relatively smooth for experts.

\subsection{Questionnaires \texorpdfstring{\&}{and} FGI}

First, we analyzed the CSRS based on the groups below (see Table \ref{Table1}). Table \ref{Table1} indicates descriptive values of communication skills from the CSRS. Experts perceived when they were paired with an expert, the communication was better than when they were paired with a student, while students felt better communication when their partner was an expert than a student. We also calculated Cronbach's alpha to assess summed rating scales' internal consistency or reliability ~\cite{cronbach1951coefficient}. In general, it is required to be at least .70, with a preferable range closer to .80 ~\cite{nunnally1978psychometric}. Cronbach's alpha for the Communication skills was .840.

\begin{table}[hbt!]
  \caption{Descriptive values of communication skills}
  \footnotesize
  \label{Table1}
  \begin{tabular}{llll} 
    \toprule
    Group & Number & Mean & SD\\
    \midrule
    Expert→Expert & 10 & 3.72 & 1.24\\
    Expert→Student & 9 & 3.52 & 1.15\\
    Student→Expert & 9 & 3.90 & 1.02\\
    Student→Student & 8 & 3.63 & 0.98\\
    \bottomrule
  \end{tabular}
\end{table}

We transcribed conversations from the FGIs based on the original data. We labeled student participants as S1, S2, .., S5 and expert participants as E1, E2, ..., E5. As a result of analyzing those FGIs, three main categories were derived: a.) Preference of Experts and Students in PP, b.) Differences in Communication Styles, and c.) Behavioral Patterns in Problem-Solving. The extracted primary insights are derived as follows: 

\subsubsection*{1. Preference of Experts and Students in PP:} 
Both experts and students recognized the value of pairing with an expert. Experts distinctly enhanced communication dynamics and confidence levels among students through active engagement and ensured task comprehension, giving confidence. This corresponds with the CSRS results, which show that participants assessed that pairing with experts had better communication than students. It also supports previous research that programmers prefer to be paired with someone with good communication and complementary skills, which are desirable in PP sessions~\cite{begel2008pair}. Conversely, when paired with students, experts felt a heightened sense of pressure and responsibility. Experts may need to encourage novices~\cite{plonka2012disengagement}. 

\begin{quote}
    \textit{E2: “Yeah, the communication skills were better when I worked with an expert.”}
    
    \textit{E1: “Maybe because I was in the expert group, I felt more responsible and nervous.”}
    
    \textit{E5: “If you're working with a student, you'll probably feel more like obligated to explain why there's a mistake there or what kind of mistake it is or something. And they tend to expect the expert to help them out. And it would be more pressure on me as an expert."}

    \textit{S2: “I felt like there was a big difference in confidence between students and experts. When comparing my experience with a student, he tried to help and throw some questions at me, and it made it a better conversation...”}

    \textit{S5: “With the expert, I actually felt quite comfortable, so I asked a lot of questions because I wanted to understand what was happening."}
\end{quote}

\subsubsection*{2. Differences in Communication Styles:}
Students noted a distinct communication style when paired with experts versus students; with experts, the interaction was more instructional, resembling a lecturer-student dynamic, while student pairings engaged in more discussion-oriented communication.

\begin{quote}    
    \textit{S3: “For me, the experiment I had with the expert was more like a lecturer style. And when I was working with a student, it took us around the same time to think about the errors. But it was more like a discussion instead of a lecture style.”}

    \textit{S5: “When I worked with the expert, I was definitely more silent because I was thinking, and he already had the answer. But when I was working with a student, I had to speak up more to encourage my partner. And I was more the lecturer."}
\end{quote}

\subsubsection*{3. Behavioral Patterns in Problem-Solving:} 

When facing challenges, participants prioritized solving the tasks over communication, leading to less interaction, whereas easier tasks encouraged more communication. Limited Zoom screen led participants to focus more on the codes than on Zoom interaction. 

\begin{quote}
\textit{E5: “If it's something easy, then I do tend to talk more because it's much easier to notice mistakes while talking. But if it’s difficult, I would, like, just completely filter out the other person while I’m trying to solve it.”}

\textit{S2: “I was just focused on the code."} 

\textit{S3: “First of all, I had to focus on the codes. So, most of the time, I wasn't really looking at the small window on the screen of my partner."}
\end{quote}

\label{conclusions}
\section{Conclusion and Future Work}

In this study, given the small sample size posing challenges in validating data from eye-tracking and questionnaires, we adopted a triangulation approach, combining FGIs to overcome these limitations and address sample size constraints, facilitating cross-validation and deeper insights into the user experience. We analyzed eye-tracking data from PP sessions to explore the relationship between eye movements and changes in communication skills across different positions, roles, and groups. Our findings highlight several key observations: participants prioritize code exploration over communication, students and experts exhibit distinct communication styles favoring the coding task, students prefer pairing with experts while experts prefer peers, and students feel more comfortable than experts, who experience greater pressure during error detection. Our future work will comprehensively analyze communication patterns, including interactions between students and chatbots, in accordance with AI development to contribute to the CS education field.



\bibliographystyle{ACM-Reference-Format}
\bibliography{references}

\end{document}